\documentclass[acmsmall,nonacm]{acmart}
\usepackage{graphicx} 
\usepackage{appendix}
\usepackage{fancyvrb}
\usepackage{algorithm}
\usepackage{algpseudocode}
\usepackage[]{mdframed}
\usepackage{enumerate}
\usepackage{hyperref}

\usepackage{listings}
\usepackage{color}
\definecolor{mygreen}{rgb}{0,0.6,0}
\definecolor{mygray}{rgb}{0.5,0.5,0.5}
\definecolor{mymauve}{rgb}{0.58,0,0.82}
\definecolor{myred}{rgb}{0.7,0.2,0.2}
\usepackage[scaled=0.85]{beramono}            
\lstdefinestyle{latexstyle}{
  language=C,
  numbers=none,
  tabsize=4,
  showspaces=false,
  showstringspaces=false
}

\usepackage[bb=boondox,bbscaled=.95,cal=boondoxo]{mathalfa}
\hypersetup{citecolor=blue}
\DeclareSymbolFontAlphabet{\mathcalorig}   {symbols}
\algrenewcommand\algorithmicrequire{\textbf{Input:}}
\algrenewcommand\algorithmicensure{\textbf{Output:}}

\title{Finding Inductive Loop Invariants using Large Language Models}
\author{Adharsh Kamath}
\affiliation{         
  \institution{Microsoft Research}  
  \country{India}
}
\email{t-adkamath@microsoft.com}

\author{Aditya Senthilnathan}
\authornote{Work was done while the author was at Microsoft Research}
\affiliation{         
  \institution{Cornell}  
  \country{USA}
}
\email{as3742@cornell.edu}

\author{Saikat Chakraborty}
\affiliation{         
  \institution{Microsoft Research}  
  \country{USA}
}
\email{saikatc@microsoft.com}

\author{Pantazis Deligiannis}
\affiliation{         
  \institution{Microsoft Research}  
  \country{USA}
}
\email{pdeligia@microsoft.com}

\author{Shuvendu K. Lahiri}
\affiliation{         
  \institution{Microsoft Research}  
  \country{USA}
}
\email{shuvendu@microsoft.com}

\author{Akash Lal}
\affiliation{         
  \institution{Microsoft Research}  
  \country{India}
}
\email{akashl@microsoft.com}

\author{Aseem Rastogi}
\affiliation{         
  \institution{Microsoft Research}  
  \country{India}
}
\email{aseemr@microsoft.com}

\author{Subhajit Roy}
\affiliation{         
  \institution{IIT Kanpur}  
  \country{India}
}
\email{subhajit@iitk.ac.in}

\author{Rahul Sharma}
\affiliation{         
  \institution{Microsoft Research}  
  \country{India}
}
\email{rahsha@microsoft.com}

\begin{document}

\newcommand{\shuvendu}[1]{{\bf SKL:{#1}}}
\newcommand{\akash}[1]{{\bf AL:{#1}}}
\newcommand{\aseem}[1]{{\bf AR:{#1}}}
\newcommand{\adharsh}[1]{{\bf AK: {#1}}}
\newcommand{\subhajit}[1]{{\bf SR: {#1}}}

\newcommand{\Omit}[1]{}

%
%
\begin{abstract}
Loop invariants are fundamental to reasoning about programs with loops. They establish properties about a given loop's behavior. When they additionally are \textit{inductive}, they become useful for the task of formal verification that seeks to establish strong mathematical guarantees about program's runtime behavior. The inductiveness ensures that the invariants can be checked locally without consulting the entire program, thus are indispensable artifacts in a formal proof of correctness. 

Finding inductive loop invariants is an undecidable problem, and despite a long history of research towards practical solutions, it remains far from a solved problem. This paper investigates the capabilities of the Large Language Models (LLMs) in offering a new solution towards this old, yet important problem. 

To that end, we first curate a dataset of verification problems on programs with loops. Next, we design a prompt for exploiting LLMs, obtaining inductive loop invariants, that are checked for correctness using sound symbolic tools. Finally, we explore the effectiveness of using an efficient combination of a symbolic tool and an LLM on our dataset, and compare it against a purely symbolic baseline.  Our results demonstrate that LLMs can help improve the state-of-the-art in automated program verification.
\end{abstract}

\maketitle

%
%
\section{Introduction}
Automatically synthesizing loop invariants is one of the classical problems in software verification. A variety of techniques have been developed to address it over multiple decades, each with their own strengths and weaknesses. With the advent of Large Language Models (LLMs), we study how well LLMs perform on this problem. To that end, we collect a dataset of programs in the C language and build a toolchain that can evaluate the success rate of a given LLM on the dataset.

Loop invariant is a logical proposition that is valid at the head of a given loop, irrespective of how many times the loop has already, or will execute. Inductiveness of loop invariants further requires that one can prove its correctness just by looking at the loop body, not the rest of the program; further the invariant is strong enough to establish the property of interest.
More formally, the problem of inductive loop invariant inference can be described as follows. Consider a \textit{Hoare triple} \cite{hoare1969axiomatic} over a loop $\{P\}\mathit{while}\ B\ \mathit{do}\ S\{Q\}$, which states that $P$ is a precondition that holds before the loop, and $Q$ is the postcondition that holds after the loop (on terminating executions). Proving this Hoare triple is valid requires finding an inductive loop invariant, i.e., an expression $\mathcal{I}$ such that the following three conditions hold. First, the loop invariant must be implied by the precondition ($P\Rightarrow \mathcal{I}$). Second, given that $\mathcal{I}$ holds at the head of the loop, and the loop executes for one arbitrary iteration, then $\mathcal{I}$ must continue to hold at the end of the loop iteration ($\{\mathcal{I}\wedge B\}S\{\mathcal{I}\}$). And finally, when the loop exits, then the loop invariant must imply $Q$ ($\mathcal{I}\wedge\neg B\Rightarrow Q$). For the sake of simplicity, we assume that the qualifier ``inductive'' is assumed and refer to such expressions $\mathcal{I}$ as loop invariants. 

\paragraph {Example.} Consider the example of a Hoare-triple for a loop in a C-like language where the variables are unbounded integers that do not overflow:
$$\{n \geq 0 \wedge x == n \wedge y == 0\} \mathit{while } \ (x >0) \ \mathit{do} \  x--; y++; \{y == n\}$$
The goal is to prove the postcondition $y == n$ on exit from the loop, when the loop is executed starting with a non-negative value $n$ in variable $x$. The invariant $\mathcal{I} \doteq \left(x + y == n \wedge x \geq 0\right)$ is inductive because the following conditions hold:
\begin{enumerate}
    \item the precondition implies the invariant, namely $$\left(n \geq 0 \wedge x == n \wedge y == 0\right) \implies \left(x+y == n \wedge x \geq 0 \right)$$
    \item the invariant is preserved by the loop body, namely $$\{\left(x + y == n \wedge x \geq 0\right)  \wedge (x > 0)\} x--; y++ \{\left(x + y == n \wedge x \geq 0 \right) \}$$
    \item the invariant implies the postcondition on exit, namely $$\left(\left(x + y == n \wedge x \geq 0 \right) \wedge \neg (x > 0) \right) \implies y == n$$
\end{enumerate}
Further, the three checks above can be efficiently expressed in a decidable logic and verified using Satisfiability-Modulo-Theories (SMT)~\cite{BarFT-SMTLIB} solvers. 

In contrast, the problem of finding loop invariants is undecidable in general. 
Earlier works have resorted to several heuristics. 
Symbolic techniques include counter-example guided abstraction refinement \cite{DBLP:journals/jacm/ClarkeGJLV03,ball-pldi01}, Craig interpolation \cite{mcmillan-cav03}, abstract interpretation with various abstract domains \cite{cousot-77,blanchet-pldi03}, syntax-guided synthesis \cite{alur2013syntax}, among others. Machine-learning-based techniques have been applied as well, including classical learning over data generated from program executions \cite{loopinvgen_padhi,DBLP:conf/sas/BrockschmidtCKK17}, active learning over decision trees \cite{garg-popl16}, reinforcement learning over graph neural networks \cite{si2018learning}, and the use of continuous logic networks \cite{yao2020learning,ryancln2inv}. Tools based on these techniques compete in the invariant inference tracks of competitions like the software verification competition (SV-COMP) \cite{svcomp} and the synthesis competition (SyGuS) \cite{sygus}. 

LLMs are now being routinely used to assist software developers \cite{copilot}, which leads to the natural question of whether they can assist in annotating loop invariants as well. We use LLMs to generate candidate loop invariants that can then be checked by a symbolic verification tool like FRAMA-C \cite{DBLP:journals/fac/KirchnerKPSY15} or Boogie \cite{barnett05}. These verifiers take a program annotated with loop invariants as input and check the validity of the Hoare triple. If the LLM generates a wrong candidate then the candidate is rejected. But if the checker passes then it has successfully demonstrated the validity of the Hoare triple.

\paragraph{Contributions} We have curated a dataset of loop invariant inference problems from several prior works on invariant inference, with the goal of evaluating LLMs for the task of loop invariant generation. We have also built a toolchain that evaluates a given LLM on this dataset, both for invariant generation as well as repair. We evaluate multiple models (GPT-4 \cite{DBLP:journals/corr/abs-2303-08774}, GPT-3.5 \cite{gpt35}, and Code Llama \cite{DBLP:journals/corr/abs-2307-09288}) and compare their performance against a state-of-the-art symbolic baseline. We further contribute a novel combination of LLMs with symbolic techniques that help tolerate inaccuracies in the model's output, only relying on the model to produce ingredients of the loop invariant rather than the whole invariant. This combination only requires a linear number of calls to the underlying symbolic solver (linear in the number of LLM generations).

In the rest of the paper, we describe this dataset (Section 2), technical details (Section 3), evaluation (Section 4), and related work (Section 5).

%
%
\section{Dataset}
We collect benchmarks from the sources listed in LoopInvGen \cite{loopinvgen_padhi}. These include all the benchmarks from prior works - Code2Inv \cite{code2Inv}, Accelerating Invariant Generation \cite{accelerating_invariant_gen}, and LinearArbitrary-SeaHorn \cite{linear_arbitrary_seahorn}. We also collect benchmarks from SV-COMP \cite{SVCOMP23}, specifically, its sub-directories starting with the prefix "loop". In total, this combined set contains 1166 benchmarks. Each benchmark is a program in the C language, where the property to be verified is present implictly in the form of assertions in the program itself. Successful verification of a program guarantees that every assertion in the program is valid in all executions of the program.

We restrict to integer programs by filtering out benchmarks that have arrays or pointers. This keeps the focus on basic mathematical reasoning, without bringing in concerns of modeling heap semantics in C. We also filter out programs with ambiguous semantics.\footnote{Some programs had the pattern \texttt{if(assert(expression))} with no clear indication of what \texttt{assert} is supposed to return.}
We also removed programs that are greater than 500 lines of code to account for current prompt size limitations in LLMs. This leaves us with 1005 benchmarks. We categorize these benchmarks based on the number of loops and methods, as shown in Table \ref{tab:benchmark_features}.

\begin{table}[h]
\centering
\begin{tabular}{|cc|c|c|}
    \cline{1-4}
      & Number of & \multicolumn{2}{|c|}{Methods}  \\ 
    \cline{3-4}

      & benchmarks & Exactly one & More than one  \\ 
    \hline
      & \multicolumn{1}{|c|}{None} & 58 & 122   \\
\cline{2-4}
        Loops & \multicolumn{1}{|c|}{Exactly one} & \textbf{555} & 30    \\
\cline{2-4}
                 & \multicolumn{1}{|c|}{More than one} & 214 & 26   \\
    \hline
\end{tabular}
\caption{Benchmark features}
\label{tab:benchmark_features}
\end{table}

Of these categories, we focus on the 555 benchmarks with exactly one loop and one method. Out of these 555 benchmarks, 469 are positive instances, i.e., there exists an invariant that is sufficient to verify all the assertions in the benchmark and 86 benchmarks are negative instances, i.e., there exists a counterexample that violates some assertion. The ground truth for each benchmark was obtained using a combination of the original benchmark source, Ultimate Automizer \cite{automizer} and manual inspection.

We make modifications to the 469 benchmarks to remove comments, modify calls to verifier-specific assert functions to ACSL assert annotations, etc., and create the final dataset used in the experiments. (ACSL \cite{e_acsl_user_manual} is the syntax that Frama-C uses for writing logical specifications.) These modifications do not alter the semantics of the programs in any way with respect to the property being verified. A detailed list of these modifications can be found in Appendix \ref{modifications}.

%
%
\section{Technical details}
Our toolchain has two main components, a large language model, $\mathcalorig{L}$, and an oracle $\mathcalorig{O}$. The input to $\mathcalorig{L}$ is text that consists of two parts: (1) Prompt template $\mathcalorig{M}$ that contains instructions relevant to finding loop invariants, common for all programs and (2) the target program $\mathcalorig{P}$, with a property to be verified in the form of assertions present in the program text.
The output of $\mathcalorig{L}$ is a set of candidate loop invariants $\mathcalorig{I}$ in the format specified in $\mathcalorig{M}$. We may query $\mathcalorig{L}$ to generate multiple completions with a non-zero sampling temperature, to generate different sets of candidate invariants.  

The oracle $\mathcalorig{O}$ checks if a given set of candidate invariants are inductive and sufficient for verifying the program. The input to $\mathcalorig{O}$ is $\mathcalorig{P}$ annotated with a given set of candidate invariants $\mathcalorig{I}$; we refer to this annotated program as $\mathcalorig{A}$($\mathcalorig{P}$, $\mathcalorig{I}$). The actual implementation of $\mathcalorig{A}$ depends on the particular formatting requirements of the orcale $\mathcalorig{O}$.

The output of $\mathcalorig{O}$ is a tuple consisting of the following values:
\begin{enumerate}
    \item A Boolean value $S$. If $S$ is true, then all the supplied candidate invariants were correct invariants of the program and were sufficient to prove correctness of the assertions as well.
    \item A string value $SyntaxError$ indicating if the oracle encountered a syntax error while parsing some candidate invariant expression. This string is necessarily empty if $S$ was true. When non-empty, $SyntaxError$ indicates the location (i.e., the candidate expression) that caused the syntax error to happen.     
    \item A set ${I}_L$ that is empty when $S$ is true or if there was a syntax error. Otherwise, it is a subset of the supplied candidate invariants. The expectation on the oracle is that whenever it fails to prove inductiveness of a set of candidate invariants, it must indicate the reason by putting blame on at least one of the candidates. This property will be made more precise later in this section.
\end{enumerate}

We describe our toolchain in the next three subsections, outlining the overall technique (Section~\ref{subsec:loopy}), and two components Houdini (Section~\ref{subsec:houdini}) and Repair (Section~\ref{subsec:repair}).

\subsection{Loopy}
\label{subsec:loopy}
Our technique---called Loopy---is detailed in Algorithm \ref{alg:loopy}. The loop on line \ref{line:LoopyLoop} iterates as many times as the user-provided bound $\mathcalorig{N}_s$ on the number of completions. For each completion, it checks if the LLM-returned set of candidate invariants are correct. If so, the entire procedure completes.  

Even when none of the completions individually were a correct set of invariants, it could happen that the necessary correct invariants were spread \textit{across} the multiple sets. To deal with such cases, Loopy computes $\mathcalorig{I}_u$ as the union of all the candidate invariant sets that were generated across multiple completions. Next, it checks if there exists a \textit{subset} $\mathcalorig{I}_s \subseteq \mathcalorig{I}_u$ that is inductive and sufficient for verifying the program. While there may be an exponential number of possible subsets of $\mathcalorig{I}_u$, it turns out that one can do this check with only a linear number of calls to the oracle (linear in the size of $\mathcalorig{I}_u$), using the Houdini algorithm \cite{houdini}. 

If Houdini is not successful, Loopy calls a procedure to repair the existing set of all LLM-generated invariants $\mathcalorig{I}_u$ in line~\ref{line:RepairCall}. 

 %
 %
\begin{algorithm}
\caption{Loopy procedure}\label{alg:loopy}
\begin{algorithmic}[1]
\Require Program $\mathcalorig{P}$, Prompt $\mathcalorig{M}$, number of times to sample invariants, $\mathcalorig{N}_s$, and number of repair retries, $\mathcalorig{N}_r$
\Ensure \textit{True} iff all the assertions in $\mathcalorig{P}$ are verified, \textit{False} otherwise
\Ensure Set of inductive invariants when the program was verified
\Procedure{Loopy} {$\mathcalorig{P}, \mathcalorig{M}, \mathcalorig{N}_s, \mathcalorig{N}_r$}
    \State $  \mathcalorig{I}_u \gets \{\} $
    \While {$ \mathcalorig{N}_s > 0 $} \label{line:LoopyLoop}
        \State $ \mathcalorig{I} \gets \mathcalorig{L}(\mathcalorig{M}, \mathcalorig{P})$ \Comment{Prompt the LLM}
        \State $  S, \_, \_ \gets \mathcalorig{O}(\mathcalorig{A}(\mathcalorig{P}, \mathcalorig{I})) $ \Comment{Annotate the program and check}
        \If {$S$ = True}
            \State \Return True, $ \mathcalorig{I} $
        \Else
            \State $ \mathcalorig{N}_s \gets \mathcalorig{N}_s - 1$
            \State $ \mathcalorig{I}_u \gets \mathcalorig{I}_u \cup \mathcalorig{I} $
        \EndIf
    \EndWhile
    \State $ S, \mathcalorig{I} \gets \Call{Houdini}{\mathcalorig{P}, \mathcalorig{I}_u} $ \Comment{Algorithm \ref{alg:houdini}}

    \If {$S$ = True}
        \State \Return True, $ \mathcalorig{I} $ \label{line:LoopyHoudiniDone}
    \Else
        \State $ S, \mathcalorig{I}  \gets \Call{Repair}{\mathcalorig{P}, \mathcalorig{I}_u, \mathcalorig{N}_r} $ \Comment{Algorithm \ref{alg:repair}} \label{line:RepairCall}
        \State \Return $ S, \mathcalorig{I}  $ 
    \EndIf

\EndProcedure
\end{algorithmic}
\end{algorithm}

\subsection{Houdini}
\label{subsec:houdini}
We describe an adaptation of Houdini to our setting in Algorithm \ref{alg:houdini}. This procedure takes as input the program $\mathcalorig{P}$ and a set of invariants $\mathcalorig{I}$ and returns $\mathit{True}$ if there exists a subset  of $\mathcalorig{I}$ that is inductive and sufficient for verification, and $\mathit{False}$ otherwise. 

Houdini works by repeatedly querying the oracle with its current set of candidate invariants $\mathcalorig{I}$. In each iteration, it either finds that $\mathcalorig{I}$ is sufficient, in which case it is done and returns True (line \ref{line:HoudiniDone}). Otherwise, it removes one or more candidates from the set $\mathcalorig{I}$ and repeats the process. Pruning happens in one of two ways. If an expression causes a syntax error, then it is pruned away (line \ref{line:HoudiniSyntaxError}). If there are no syntax errors, then the expressions in $I_L$ are pruned away (line \ref{line:HoudiniPrune}). If $I_L$ was empty, then the procedure returns False; this indicates the case when the current set $\mathcalorig{I}$ is a valid inductive invariant, but still not sufficient (i.e., strong enough) to prove the program correct. 

A candidate might not be inductive for one of two reasons. It might not hold on entry to the loop (i.e., not implied by the precondition), or that it may not be preserved by the loop body. The complication is that preservation of one candidate might rely on other candidates. Thus, we require that for  each candidate $i$ that is included in $I_L$ by the oracle, either $i$ did not hold at entry to the loop, or that $i$ was not preserved even when assuming \textit{all} candidate invariants (even failed ones) to hold at the beginning of the loop iteration. There must be at least one such candidate if the candidate set is not collectively inductive. In other words, there is no support for candidates in $I_L$ to be inductive invariants, whereas the rest of the candidates might still be proved inductive. Checkers such as Frama-C \cite{DBLP:journals/fac/KirchnerKPSY15} and Boogie \cite{barnett05} are able to provide such information. However, we do note that this property is only required for optimality of Houdini. Irrespective of how $I_L$ is populated, the guarantee that Algorithm \ref{alg:houdini} returns an inductive set of invariants when it returns \textit{true} is a valid guarantee. It is just that Algorithm \ref{alg:houdini} might miss finding a valid subset among the candidates even when one exists. 

Furthermore, note that Houdini makes a linear number of calls to the Oracle -- at most as many calls as the size of the input candidate invariant set.

%
%
\begin{algorithm}
\caption{Houdini procedure}\label{alg:houdini}
\begin{algorithmic}[1]
\Require Program $\mathcalorig{P}$ and a set of invariants $\mathcalorig{I}$
\Ensure (\textit{True}, $\mathcalorig{I}_0$) if there exists $\mathcalorig{I}_0 \subseteq \mathcalorig{I}$ such that $\mathcalorig{O}(\mathcalorig{A} (\mathcalorig{P}, \mathcalorig{I}_0))$ is \textit{True}. (\textit{False},$\{\}$) otherwise
\Procedure{Houdini} {$\mathcalorig{P}$, $\mathcalorig{I}$}
\While {$ \lvert \mathcalorig{I} \lvert \: > 0 $}
    \State $ S, I_L, SyntaxError \gets \mathcalorig{O}(\mathcalorig{A} (\mathcalorig{P}, \mathcalorig{I}))$) \Comment{Annotate the program and check}
    \If {$S = True$}
        \State \Return (True, $\mathcalorig{I}$) \label{line:HoudiniDone}
    \EndIf
    \If {$SyntaxError$ = ``error while parsing invariant $i$''}
        \State $ \mathcalorig{I} \gets \mathcalorig{I} - \{i\} $ \label{line:HoudiniSyntaxError}
    \Else
         \If {$ I_L == \{\} $}
            \State break
         \Else
            \State $\mathcalorig{I} \gets \mathcalorig{I} - I_L $  \label{line:HoudiniPrune}
         \EndIf
    \EndIf
\EndWhile
\State \Return (False, $\{\}$)
\EndProcedure
\end{algorithmic}
\end{algorithm}

\subsection{Repair}
\label{subsec:repair}
The Repair procedure takes a set of invariants $\mathcalorig{I}$ and attempts to repeatedly generate a new set of invariants $\mathcalorig{I}'$ that are "close" to $\mathcalorig{I}$ but contains an inductive subset. 
It either returns (\textit{True}, $\mathcalorig{I}'$) where $\mathcalorig{I}'$ is inductive, or returns 
(\textit{False},$\{\}$).

The procedure, shown as Algorithm \ref{alg:repair}, repeatedly calls the oracle $\mathcalorig{O}$ on a set of invariants $\mathcalorig{I}$ (lines~\ref{line:InitialRepairOracleQuery} and ~\ref{line:RepairOracleQuery}) and then queries the LLM $\mathcalorig{L}$ to repair the set of invariants using a repair prompt $\mathcalorig{M}_r$ over the error information present in $I_L$ and $SyntaxError$ (line~\ref{line:RepairLLMPrompt}).
Details of the prompt are present in Appendix~\ref{prompts_and_parameters}.
The new set of invariants is queried for inductiveness (line~\ref{line:RepairOracleQuery}) or passed to Houdini procedure to find an inductive subset (line~\ref{line:RepairHoudiniQuery}).
The process is repeated for  $\mathcalorig{N}_r$ times, after which the procedure returns (\textit{False},$\{\}$) to indicate failure. 

%
%
\begin{algorithm}
\caption{Repair procedure}\label{alg:repair}
\begin{algorithmic}[1]
\Require Program $\mathcalorig{P}$, a set of invariants $\mathcalorig{I}$, and number of repair retries $\mathcalorig{N}_r$
\Ensure (\textit{True}, $\mathcalorig{I}'$) if $\mathcalorig{O}(\mathcalorig{A} (\mathcalorig{P}, \mathcalorig{I}'))$ is \textit{True}. (\textit{False},$\{\}$) otherwise
\Procedure{repair} {$\mathcalorig{P}, \mathcalorig{I}, \mathcalorig{N}_r$}
\State $ S, I_L, SyntaxError \gets \mathcalorig{O}(\mathcalorig{A}(\mathcalorig{P}, \mathcalorig{I})) $ \label{line:InitialRepairOracleQuery}
\State $ n \gets 0 $
\While {$ n < \mathcalorig{N}_r $}
\State $ \mathcalorig{I} \gets \mathcalorig{L}(\mathcalorig{M}_r(I_L, SyntaxError), \: \mathcalorig{A}(\mathcalorig{P}, \mathcalorig{I}) )$ \Comment{Instruct the LLM to repair} \label{line:RepairLLMPrompt}
\State $  S, I_L, SyntaxError \gets \mathcalorig{O}(\mathcalorig{A}(\mathcalorig{P}, \mathcalorig{I})) $ \Comment{Use the new invariants to verify} \label{line:RepairOracleQuery}
    \If {$S = True$}
        \State \Return $(True, \mathcalorig{I})$
    \Else
    \State $ S, \mathcalorig{I}' \gets \Call{Houdini}{\mathcalorig{P}, \mathcalorig{I}} $ \Comment{Algorithm \ref{alg:houdini}} \label{line:RepairHoudiniQuery}
        \If {$ S = True $}
            \State \Return $(True, \mathcalorig{I}')$
        \Else
            \State $n \gets n + 1$
        \EndIf
    \EndIf
\EndWhile
\State \Return $(False, \{\})$
\EndProcedure
\end{algorithmic}
\end{algorithm}

%
%
\section{Evaluation}
We design our experiments with the goal of answering the following research questions. We call a candidate set of invariants $\mathcalorig{I}$ as \textit{correct} for a program $\mathcalorig{P}$ when $\mathcalorig{O}(\mathcalorig{A} (\mathcalorig{P}, \mathcalorig{I}))$ returns true. 
\begin{enumerate}
    \item[RQ1] How often can LLMs generate a correct set of loop invariants for a C program?
    \item[RQ2] How often can LLMs generate elements of a correct set of loop invariants needed to verify a C program?
    \item[RQ3] How do different foundational models compare in their ability to find inductive invariants? 
    \item[RQ4] How often can LLMs fix incorrect invariants using error messages from the oracle?
    \item[RQ5] What are the program features where LLMs fail to generate correct invariants? 
    \item[RQ6] How does the performance of Loopy compare with state-of-the-art symbolic verifiers? 
\end{enumerate}

\paragraph{Experimental setup}
To answer these RQs, we instantiate the previously described toolchain with Frama-C as the oracle $\mathcalorig{O}$, and GPT-4, GPT-3.5-Turbo, and CodeLlama-34b-Instruct as the Large language model $\mathcalorig{L}$. Frama-C was configured to use only the \textit{WP} plugin for verification, which is suited to check ACSL annotations in C programs. Note that Frama-C does not attempt constructing inductive invariant on its own, rather it is focused on verifying the correctness of any candidate invariants as well as assertions in the program that it is given as input. Additionally, the \textit{WP} plugin was configured to use Z3 \cite{z3}, Alt-Ergo \cite{alt-ergo}, and CVC4 \cite{cvc4} as the external provers, with a timeout of 3 seconds.

We use a symbolic tool Ultimate Automizer \cite{automizer} to establish a symbolic baseline performance. Ultimate's goal is to either establish invariants that prove program correctness, or return a counterexample. It provides a strong baseline comparison because it has been competing in the software verification competition SV-COMP \cite{svcomp}, which largely contains a superset of our dataset, for a number of years and winning in several categories. 
Ultimate was configured with a core-toolchain timeout of 300 seconds. Commonly used tool configurations are included with releases of Ultimate Automizer. Of these configurations, we use \textit{AutomizerReach.xml} as the toolchain file, and \textit{svcomp-Reach-64bit-Automizer\_Default.epf} as the settings file.

LLM hyper-parameters that we used are given in Appendix \ref{prompts_and_parameters}. 

\begin{figure}[htb]
\includegraphics[scale=0.6]{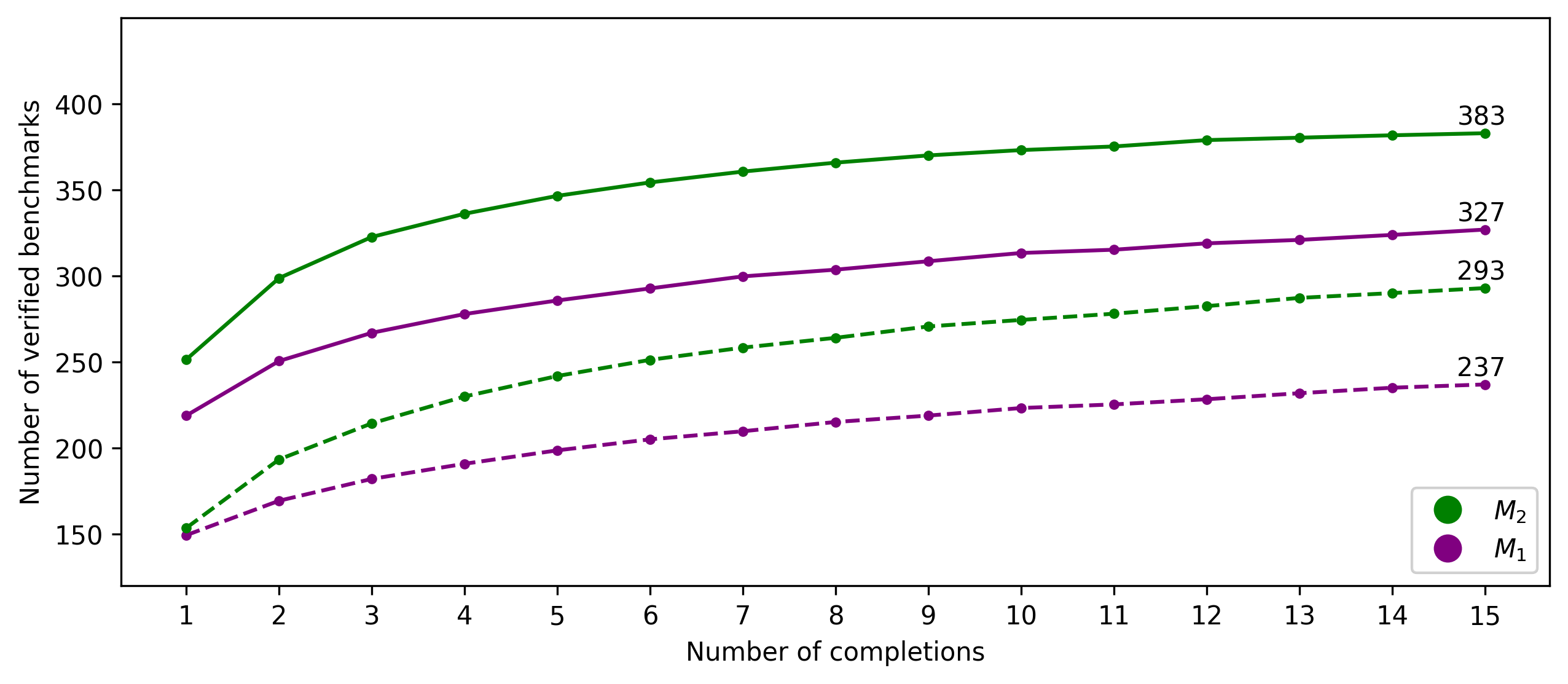}
\caption{Success rate of Loopy with GPT-4 as the number of completions is varied. Dashed lines depict the performance of Loopy without Houdini and without Repair, for different prompts. Solid lines depict the performance of Loopy with Houdini, but without Repair.}
\label{fig:pass_at_k}
\centering
\end{figure}

\subsection{[RQ1] Using LLMs to generate loop invariants}
To answer RQ1, we measure the number of benchmarks for which the Loopy procedure (Algorithm \ref{alg:loopy}) successfully verifies the benchmark program without calling either Houdini or the Repair procedures. 

We experimented with two different prompts. Prompt $ \mathcalorig{M}_1 $ includes the input program and simply asks for loop invariants. Prompt $ \mathcalorig{M}_2 $ additionally includes instructions that detail the concept of a loop invariant and additionally include certain hints or ``nudges'' that offer guidelines on patterns that a loop invariant might contain. These nudges are generic and not tailored to specific program instances. We include both these prompts in Appendix \ref{prompts_and_parameters}.

Results are reported in Figure \ref{fig:pass_at_k}; consider only the dashed lines for now. It shows success rate of Loopy (number of benchmarks verified) as the number of completions is varied. 
We account for the non-deterministic nature of LLMs in the standard way followed when computing the ``pass@k'' metric as described in Chen et al. \cite{codex}: we first generate the maximum number of completions ($15$) and compute its success rate. Then, the success rate for $k$ completions is obtained as the expectation over a random sample of size $k$ out of the $15$ completions.

The figure shows that multiple completions help, although there are diminishing returns after around $8$ completions. Adding instructions to the prompt helps, with $\mathcalorig{M}_2$ performing significantly better than $\mathcalorig{M}_1$. Prompt $\mathcalorig{M}_2$ helps solve $23\%$ more instances than prompt $\mathcalorig{M}_1$ (293 benchmarks for $\mathcalorig{M}_2$, compared to 237 benchmarks for $\mathcalorig{M}_1$).
 
\begin{figure}[t]
\includegraphics[scale=0.65]{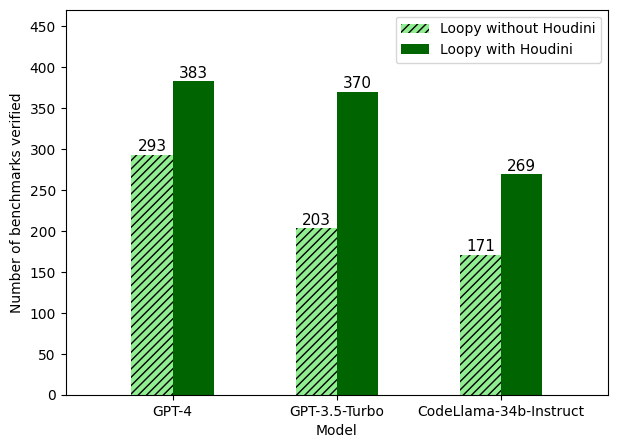}
\caption{Performance of Loopy instantiated with different LLMs, with $15$ completions and prompt $\mathcalorig{M}_2$}
\label{fig:model_comparison}
\centering
\end{figure}

\subsection{[RQ2] LLMs generating ingredients of invariant}
For RQ2, we experiment with Loopy while using Houdini but not the Repair procedure. The results are shown as the solid lines of Figure \ref{fig:pass_at_k}. Success rate for $k < 15$ is computed as follows: as an average over randomly sampling $k$ out of the $15$ completions, taking their union and running Houdini.

Houdini has a significant positive impact on the performance of Loopy. With $15$ completions and the $ \mathcalorig{M}_2$ prompt, the use of  Houdini increases success rate by \textbf{30.7}\% (from 293 to 383 solved benchmarks). Prompt $\mathcalorig{M}_2$ again does better over $\mathcalorig{M}_1$ (327 to 383 solved benchmarks). We can thus conclude from these experiments that LLMs are much better at finding ingredients of loop invariants than they are at finding the complete set of invariants. Furthermore, including instructions in the LLM prompt helps improve accuracy.

\begin{figure}[t]
\includegraphics[scale=0.4]{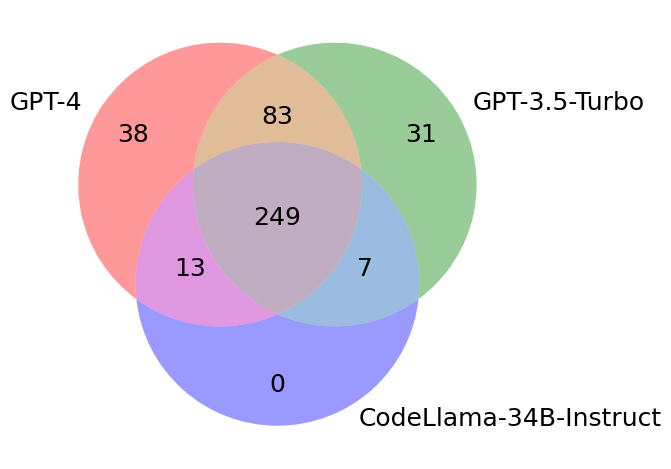}
\caption{Intersection of benchmarks verified by Loopy (with Houdini, without Repair), using different LLMs}
\label{fig:llm_intersection_union}
\centering
\end{figure}

\subsection{[RQ3] Comparing different LLMs}

Results for Loopy without Repair, as we vary the LLM are shown in Figure \ref{fig:model_comparison}. Number of completions used for this experiment was fixed at $15$. GPT-4 shows a superior performance compared to the other models, although GPT-3.5-Turbo is a close second with 370 solved benchmarks when using Houdini. Interestingly, using Houdini helps other models ``catch up'' with GPT-4 by significantly increasing their success rates. 

Figure \ref{fig:llm_intersection_union} shows the intersection among the benchmarks verified using the different LLMs. GPT-4 has the most number of exclusively-solved benchmarks. GPT-3.5-Turbo is able to solve several cases that GPT-4 could not. Using multiple LLMs can be beneficial for Loopy.

\subsection{[RQ4] Using LLMs to fix incorrect invariants}
To answer RQ4, we use the full Loopy algorithm, i.e., when using both Houdini as well as the Repair procedures. Since Loopy with GPT-4 performs the best in comparison with the other LLMs, we restrict this experiment to GPT-4. For the same reason, we pick prompt $\mathcalorig{M}_2$.

With using Repair, we must decide on the relative budget between generation and repair. Loopy can be configured with number of completions $\mathcalorig{N}_s$ used for generation and the number of iterations of the repair loop $\mathcalorig{N}_r$ (see Algorithm \ref{alg:loopy}). For a fair comparison we should have $\mathcalorig{N}_s + \mathcalorig{N}_r = 15$ so that the use of Repair does not increase the number of LLM responses for Loopy. Since the performance of Loopy-without-Repair starts to plateau at around $8$ completions (Figure \ref{fig:pass_at_k}), we set $\mathcalorig{N}_s = 8$ and $\mathcalorig{N}_r = 7$. In this case, our evaluation shows that Loopy-with-Repair verifies an additional 15 benchmarks compared to Loopy-without-Repair. This implies a total of 398 solved benchmarks.

Consider the program in Example \ref{ex:repair_success1}. To verify this program, Loopy initially generates the invariant conjuncts shown in Example \hyperref[ex:before_repair]{1.1}. The conjuncts in bold capture the behaviour of  $x$ and $y$ after the first iteration of the loop. However, there is no expression that captures the program behaviour before the loop begins (where $y$ is unconstrained). Upon invoking the Repair procedure, Loopy generates the invariant shown in Example \hyperref[ex:before_repair]{1.2} in the fourth repair try. This invariant allows $y$ to take any value at loop entry, while still capturing the loop behaviour. This invariant is inductive and sufficient for verification.

\begin{minipage}[t]{0.4\textwidth}
    \begin{lstlisting}[frame=single, columns=flexible, label={ex:repair_success1}, caption=Repairing incorrect invariants]
void main()
{
    int x = 1;
    int y;

    while (x <= 10)
    {
        y = 10 - x;
        x = x + 1;
    }

    assert(y < 10);
}
\end{lstlisting}
\end{minipage}\hfill
\begin{minipage}[t]{0.45\textwidth}
    \bigbreak
    \begin{mdframed}
    \label{ex:before_repair}
        $ (1 \leq x \leq 11) \land (y = 10 - x) \land \\ (0 \leq y \leq 9) \land (\mathbf{\pmb{y = 10 - (x - 1)}}) \land \\ (\mathbf{\pmb{y < 10}})$
    \end{mdframed}
    \smallbreak
    \centering
    Ex. 1.1. Invariant before repair

    \bigbreak
    \bigbreak

    \begin{mdframed}
    \label{ex:after_repair}
        $
            (x = 1 \lor \mathbf{\pmb{y = 10 - x + 1}}) \land \\
            (x = 1 \lor \mathbf{\pmb{y < 10}})
        $
    \end{mdframed}
    \smallbreak
    Ex. 1.2. Invariant after repair

\end{minipage}\hfill

\subsection{[RQ5] Qualitative analysis of the failure cases}
To answer RQ5, we consider the benchmarks that remain unverified even with the repair procedure. We examined these programs and manually came up with the \textit{simplest} set of loop invariants that we could find. (The notion of ``simple'' here is subjective.) Observing these invariants, we classified the failed benchmarks into five categories based on the ground truth invariant (which may not be unique.) The classification can be found in Table \ref{tab:qual_analysis}. The classification is not indicative of features that are beyond LLMs today; there are also benchmarks in each of these category that LLMs are able to solve. We do believe, however, that these feature are somewhat predictive of programs that LLMs find difficult to reason about. 

The first category of failure cases are benchmarks that require disjunctions in the invariant. These benchmarks can be described as either having loops with multiple \textit{phases} (i.e., the code path taken inside the loop body changes several times, based on flags or other branches, as the loop continues to iterate), or assertions that depend on whether the loop will be entered or not (i.e., one needs a disjunct to account for the case that the loop may not execute even once). The program in Example \ref{ex:loopy_failure1} has a multi-phase loop. It starts with $x$ and $y$ both at 0; then both increase by 1 in each iteration until $x$ reaches 50, after which $x$ continues to increase by 1 while $y$ starts to decrease. Describing these ``phases'' requires clauses, one for each phase, connected by disjunctions 

The second category of failure cases are benchmarks whose ground truth invariants requires a clause with at least three variables. An example of an invariant in this category is the following: $(0 < p) \wedge (2*ielen + leader\_len \leq bufsize\_0) \wedge (p = leader\_len + 2*i)$.

The third category of failures are benchmarks where more precise constraints were required, compared to what was generated by Loopy. For instance, Loopy finds the invariant $ (k = x + y + z) \land (x \leq y) \land (y = z)$ for the benchmark in Example \ref{ex:loopy_failure2}. While this invariant is inductive, it is not sufficient to prove the assertion. If only the second clause were more precise, claiming $x = y$, the benchmark would have been verified.

The fourth category requires reasoning about floating point arithmetic. It was hard to us, even manually, to come up with their ground-truth invariants. 

The final category of failures is due to issues in Frama-C. In these cases, Loopy came up with a correct set of invariants, but  Frama-C rejected them, likely due to imprecise reasoning inside the solver. We believe that it should be possible to strengthen Frama-C to handle these cases because the reasoning required is very much within reach of modern SMT solvers. (Some of the invariants, for instance, required reasoning with integer mod operation that is supported by solvers like Z3.)

\begin{table}[t]

\begin{tabular}{|l|l|}
\hline
 \textbf{Category} & \textbf{No. of benchmarks}  \\ \hline
 Ground truth invariant requires disjunctions & 44  \\ \hline
 Ground truth invariant requires long clauses & 5  \\ \hline
 Ground truth invariant requires more precise constraints & 9  \\ \hline
 Ground truth invariant requires floating point reasoning & 3  \\ \hline
 Frama-C failure & 10  \\ \hline
\end{tabular}
\caption{Qualitative analysis of the failed benchmarks}
\label{tab:qual_analysis}
\end{table}

\begin{minipage}[t]{0.4\textwidth}
    \begin{lstlisting}[frame=single, columns=flexible, label={ex:loopy_failure1}, caption=Requires disjunction]
int main()
{
	int x = 0, y = 0, flag = 0;
	while (flag < 1) {
		if (y < 0)
			flag = 1;

		if (flag < 1)
			x = x + 1;

		if (x < 50)
			y = y + 1;
		else
			y = y - 1;
	}
	assert(y == -2 && x == 99);
	return 0;
}
\end{lstlisting}
\end{minipage}\hfill
\begin{minipage}[t]{0.4\textwidth}
    \begin{lstlisting}[frame=single, columns=flexible, label={ex:loopy_failure2}, caption=Requires more precise constraints]
void main()
{
	int x = 0;
	int y = 0;
	int z = 0;
	int k = 0;

	while (unknown_int())
	{
		if (k % 3 == 0) 
		{
			x++;
		}
		y++;
		z++;
		k = x + y + z;
	}

	assert(x == y && y == z);
}
    \end{lstlisting}
\end{minipage}\hfill

\subsection{[RQ6] Symbolic baseline}

\begin{figure}[t]
\includegraphics[scale=0.4]{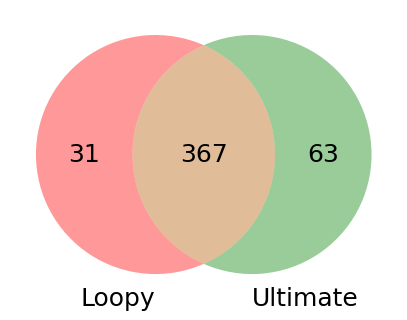}
\caption{Intersection of benchmarks verified using Loopy and Ultimate}
\label{fig:ultimate_gpt4_union}
\centering
\end{figure}

We compare the performance of Loopy (with Houdini as well as Repair) against Ultimate. Overall, Ultimate is able to solve more benchmarks (430 out of 469) compared to Loopy (398 out of 469). However, there are several benchmarks that only Loopy can solve, demonstrating that LLMs have the potential of improving state-of-the-art in program verification. Consider the program in Example \ref{ex:loopy_success_1}. The assertion in line 12 can be verified using the loop invariant $ x + y = n $ for the while loop. Loopy successfully finds this loop invariant while Ultimate fails to do so. Similarly in Example \ref{ex:loopy_success_2}, the loop invariant \: $ k \leq 1000000 \: \land \: k \leq i $ can be used to verify the program. Loopy successfully finds this, but Ultimate does not. Figure \ref{fig:ultimate_gpt4_union} shows a comparison between Loopy and Ultimate. Loopy is able to solve $31$ benchmarks on which Ultimate failed.

\begin{minipage}[t]{0.4\textwidth}
    \begin{lstlisting}[frame=single, columns=flexible, label={ex:loopy_success_1}, caption=Loopy succeeds but Ultimate fails]
int main()
{
    unsigned int n = unknown_uint();
    unsigned int x = n, y = 0;

    while (x > 0)
    {
        x--;
        y++;
    }

    assert(y == n);
    return 0;
}
\end{lstlisting}
\end{minipage}\hfill
\begin{minipage}[t]{0.45\textwidth}
\begin{lstlisting}[frame=single, columns=flexible, caption=Loopy succeeds but Ultimate fails, label={ex:loopy_success_2}]
int main()
{
    int i = 0;
    int k = 0;

    while (i < 1000000)
    {
        int j = unknown_int();
        if (!(1 <= j && j < 1000000))
            return 0;
        i = i + j;
        k++;
    }

    assert(k <= 1000000);
    return 0;
}
\end{lstlisting}
\end{minipage}\hfill

%
%
\section{Related work}

The success of LLMs in the domain of code-generation tasks \cite{copilot,copilotx,codewhisperer} has raised interest in applying them towards code reasoning tasks as well. With a focus on program verification, this naturally leads to the question of synthesizing invariants for programs. 

Pei et al. \cite{pei2023can} study this problem by building dataset of programs and corresponding invariants and then fine-tune a pre-trained LLM on this dataset. Our approach does not rely on pre-training and directly evaluates the capabilities of foundational models. Furthermore, Pei et al. do not focus on generating \textit{inductive} invariants that are necessary for establishing a formal proof of correctness.

Chakraborty et al. \cite{chakraborty2023ranking} study the problem of automatically ranking invariants that are generated by an LLM. Their approach, iRank, builds a custom model for ranking candidate invariants. iRank is orthogonal and complementary to our work; we can use it as a heuristic for decreasing the number of calls to the oracle by only checking highly-ranked invariant candidates. Their approach for candidate invariant generation does not consider Houdini and do not leverage LLMs for repairing invariants. Further, iRank  uses a lower level SMT-encoding of the input programs. We query LLMs with C programs instead, arguably a syntax that LLMs should  understand better.

%
%
\section{Conclusion}
This paper evaluates the capabilities of pre-trained Large Language Models towards unlocking the potential of formal verification. Towards this quest, we define and evaluate latest LLMs on the task of generating inductive invariants for program loops. We present a dataset and a toolchain for evaluation. Our results show that a simple and efficient combination of LLMs with formal pruning (Houdini) is able to out-perform state-of-the-art symbolic verifiers on multiple benchmarks. Our results should inspire further research in combining formal verification and LLMs.

%
%
\bibliographystyle{acm}
\bibliography{main}

%
%
\appendix
\section{Appendix}

\subsection{Prompts and parameters}
\label{prompts_and_parameters}
We set the sampling temperature to $0.7$ and max tokens in one generation to $2000$ for all queries made to the LLM. The prompts $\mathcalorig{M}_1$ and $\mathcalorig{M}_2$ are given below.

\noindent Prompt $\mathcalorig{M}_1$:
\begin{mdframed}
Consider the following C program: \\
\textasciigrave\textasciigrave\textasciigrave \\
\{\{ code \}\} \\
\textasciigrave\textasciigrave\textasciigrave \\

\noindent Output the loop invariants for the loop in the program above. \\
Output all the loop invariants in one code block. For example: \\
\textasciigrave\textasciigrave\textasciigrave \\
/*@ \\
    loop invariant i1; \\
    loop invariant i2; \\
*/ \\
\textasciigrave\textasciigrave\textasciigrave \\
\end{mdframed}

\bigbreak

\noindent Prompt $\mathcalorig{M}_2$
\begin{mdframed}
You are a helpful AI software assistant that reasons about how code behaves. 
Given a program, you can find loop invariants, which can then be used to 
verify some property in the program.  Frama-C is a software verification 
tool for C programs. The input to Frama-C is a C program file with ACSL 
(ANSI/ISO C Specification Language) annotations.
For the given program, find the necessary loop invariants of the while 
loop to help Frama-C verify the post-condition.

\noindent Instructions:
\begin{itemize}
    \item Make a note of the pre-conditions or variable assignments in the program.
    \item Analyze the loop body and make a note of the loop condition.
    \item Output loop invariants that are true 
    \begin{enumerate}[(i)]
        \item before the loop execution, 
        \item in every iteration of the loop and 
        \item after the loop termination, 
    \end{enumerate}
such that the loop invariants imply the post condition.
    \item If a loop invariant is a conjunction, split it into its parts.
    \item Output all the loop invariants in one code block. \\ For example: \\
\textasciigrave\textasciigrave\textasciigrave \\
/*@ \\
    loop invariant i1;\\
    loop invariant i2;\\
*/ \\
\textasciigrave\textasciigrave\textasciigrave
\end{itemize}
Rules:
\begin{itemize}
    \item **Do not use variables or functions that are not declared in the program.** 
    \item **Do not make any assumptions about functions whose definitions are not given.**
    \item **All undefined variables contain garbage values. Do not use variables that have garbage values.**
    \item **Do not use keywords that are not supported in ACSL annotations for loops.**
    \item **Variables that are not explicitly initialized, could have garbage values. Do not make any assumptions about such values.**
    \item **Do not use the \textbackslash at(x, Pre) notation for any variable x.**
    \item **Do not use non-deterministic function calls.** \\
\end{itemize}

\noindent Consider the following C program: \\
\textasciigrave\textasciigrave\textasciigrave \\
\{\{ code \}\} \\
\textasciigrave\textasciigrave\textasciigrave 

\noindent You are allowed to use implication to take care of the conditional nature of the code. Use implication (==>) instead of using if-then.

\noindent For all variables, add conjunctions that bound the maximum and minimum values that they can take, if such bounds exist.

\noindent If a variable is always equal to or smaller or larger than another variable, add a conjunction for their relation.

\noindent If the assertion is guarded by a condition, use the guard condition in an implication.

\noindent If certain variables are non-deterministic at the beginning or end of the loop, use an implication to make the invariant trivially true at that location. 

\noindent Output the loop invariants for the loop in the program above. Let's think step by step.
\end{mdframed}

\bigbreak

The prompt, $\mathcalorig{M}_r$, used in the Repair procedure is as follows:

\bigbreak

\begin{mdframed}
You are a helpful AI software assistant that reasons about how code behaves. Given a program, find loop invariants or fix incorrect invariants, which can then be used to verify the assertion(s) in the program. 
Frama-C is a software verification tool for C programs. The input to Frama-C is a C program file with ACSL (ANSI/ISO C Specification Language) annotations.
For the given program with loop invariants and the output from Frama-C, fix the invariants that are not inductive or have syntax errors, so that the program can be verified.

\noindent Instructions:
\begin{itemize}
    \item Make a note of the pre-conditions or variable assignments in the program.
    \item Analyze the loop body and make a note of the loop condition. 
    \item For each invariant that has syntax errors, fix the syntax errors and ensure that ACSL syntax is followed.
    \item For each invariant that is not inductive, there are four possibilities:
        \begin{itemize}
            \item The invariant is partially proven to be inductive: In this case, try to fix the non-inductive invariants that this invariant depends on.
            \item The invariant is preserved but not established: In this case, add a clause to the invariant such that the new invariant  holds before the loop execution.
            \item The invariant is established but not preserved: In this case, add a clause to the invariant such that the new invariant holds after a loop iteration if it holds before the loop iteration.
            \item The invariant is neither established nor preserved: In this case, remove it or replace it with a different inductive invariant, or add clauses to the invariant such that the new invariant is established and preserved.
        \end{itemize}
    \item Finally output loop invariants that are true 
    \begin{enumerate}[(i)]
        \item before the loop execution, 
        \item in every iteration of the loop and 
        \item after the loop termination, 
    \end{enumerate}
    such that the loop invariants can be used to prove the assertion(s) that are labeled as "Unproven" by Frama-C.
    \item If a loop invariant is a conjunction, split it into its parts.
    \item Output all the loop invariants in one code block. For example: \\ 
    \textasciigrave\textasciigrave\textasciigrave \\
    /*@ \\
        loop invariant i1; \\
        loop invariant i2; \\
    */ \\
    \textasciigrave\textasciigrave\textasciigrave \\
\end{itemize}

\noindent Rules:
\begin{itemize}
    \item **All undefined variables contain garbage values. Do not use variables that have garbage values.**
    \item **Do not make assumptions about the values returned by non-deterministic function calls.**
    \item **Do not use variables or functions that are not declared in the program.** 
    \item **Do not make any assumptions about functions whose definitions are not given.**
    \item **Do not use keywords that are not supported in ACSL annotations for loops.**
    \item **Use \textbackslash at(x, LoopEntry), instead of \textbackslash at(x, Pre) to refer to the value of a variable x just before the beginning of loop execution.**
    \item **Variables that are not explicitly initialized, could have garbage values. Do not make any assumptions about such values.**
\end{itemize}

\noindent Consider the following program with loops annotated with the respective loop invariants: \\
\textasciigrave\textasciigrave\textasciigrave \\
\{\{ code \}\} \\
\textasciigrave\textasciigrave\textasciigrave \\

\noindent Frama-C returns the following message: \\
\textasciigrave\textasciigrave\textasciigrave \\
\{\{ error \}\} \\
\textasciigrave\textasciigrave\textasciigrave \\

\noindent If the error message indicates a syntax error in the loop annotation, fix the line with the syntax error.
If the error message says "unbound logic variable x", it means that the variable x is not accessible at the location specified in the loop annotation.

\noindent If an invariant is partially proven to be inductive, it means that the verifier is able to prove that the invariant is inductive, under the assumption that the other invariants are inductive. But some of the other invariants are not inductive. \\
If an invariant is established, it means that the verifier is able to prove that the invariant holds before the loop. \\
If an invariant is preserved, it means that the verifier is able to prove that the invariant holds after the loop, assuming that it holds before the loop.

\noindent To fix the non-inductive invariants, try the following:
If an invariant is partially proven to be inductive, make sure it is actually inductive. If it is not, remove it or replace it with a different inductive invariant. If it is, then fix the other non-inductive invariants. \\
If an invariant is preserved but not established, add a clause to the invariant to make it established (a clause that makes the invariant hold before the loop begins). \\
If an invariant is established but not preserved, add a clause to the invariant to make it preserved (a clause that makes the invariant hold after the loop ends, assuming that it holds before the loop begins). \\
If an invariant is neither established nor preserved, remove it or replace it with a different inductive invariant.

\noindent If none of the above is possible, add a new loop invariant to strengthen the existing invariants.

\noindent Please fix all the above-mentioned errors, and provide the corrected loop invariants for the program that are sufficient to verify the assertions in the program.
Use the ACSL syntax to specify the invariants. Output only one code block in the end, with the fixed invariants. **Do not output any other code blocks**.
Let's think step by step.
\end{mdframed}

\subsection{Modifications}
\label{modifications}

The important modifications we make to the benchmarks to create our dataset are as follows:

\begin{itemize}
    \item Remove all comments - We noticed that some benchmarks had comments about the loop or the invariant required for verification. To maintain uniformity, we removed all comments from the benchmarks.
    \item Replace \texttt{\_\_VERIFIER\_*} functions with common functions. Different benchmark sources use different conventions while referring to function calls with non-deterministic return values, or calls to their version of an ``assume'' function. To maintain uniformity, we replace these calls with calls to \texttt{unknown\_int()}, \texttt{unknown\_float()}, etc. for the corresponding non-deterministic function calls, and use a macro to define \texttt{assume}.
    \item Convert \texttt{\_\_VERIFIER\_assert()} calls to ACSL annotation. Since we use Frama-C as our oracle, we convert the original assertions in the benchmark to ACSL annotations, which are comments that are immediately followed by '@'. For example, \texttt{//@ assert()}.
    \item Convert Error label statements and \texttt{reach\_error()} calls to ACSL. Some benchmarks contain an error-labeled statement or a \texttt{reach\_error()} function call, with the intention that it should be unreachable in the program. In such cases, we replace the said statements with \texttt{//@ assert (\textbackslash false)}.
    \item Fix \texttt{main} function. In some benchmarks, the \texttt{main} function does not return a value whose type matches the function's signature. Either a different value is returned or no value at all. This might be acceptable to most standard C compilers because they assign some default behaviour in such cases, but Frama-C returns an error. In order to avoid this, we add a default return statement that matches the return type of the function. This does not affect the verification task.
    \item Pre-processor lines. Some benchmarks might contain lines that are the result of the program file being passed through a pre-processor. We discard such lines since they are of no significant meaning for the verification task.
\end{itemize}

\end{document}